\documentclass[%
 reprint,
 amsmath,amssymb,
 aps,
 pra,
floatfix,longbibliography
]{revtex4-1}
\usepackage{graphicx}
\usepackage{dcolumn}
\usepackage{bm}
\usepackage{float}

\usepackage[utf8]{inputenc}
\usepackage[T1]{fontenc}
\usepackage{mathptmx}
\usepackage{etoolbox}
\usepackage{svg}
\makeatletter
\def\@email#1#2{%
 \endgroup
 \patchcmd{\titleblock@produce}
  {\frontmatter@RRAPformat}
  {\frontmatter@RRAPformat{\produce@RRAP{*#1\href{mailto:#2}{#2}}}\frontmatter@RRAPformat}
  {}{}
}%
\makeatother
\begin{document}

\preprint{AIP/123-QED}

\title{Single-Shot Ionization-Based Transverse Profile Monitor for Pulsed Electron Beams}


\author{P. Denham\textsuperscript{1,*}} 
\author{A. Ody\textsuperscript{1}}
\author{P. Musumeci\textsuperscript{1}}
\author{N. Burger\textsuperscript{2}}
\author{N. Cook\textsuperscript{3}}
\author{G. Andonian\textsuperscript{1,2}}

\affiliation{%
  \textsuperscript{1}University of California Los Angeles, Los Angeles, CA, USA
}

\affiliation{%
  \textsuperscript{2}RadiaBeam Technologies, Santa Monica, CA, USA
}

\affiliation{%
  \textsuperscript{3}RadiaSoft LLC, Boulder, CO, USA
}

\thanks{Contact author: \text{pdenham@physics.ucla.edu}}
\date{\today}

\begin{abstract}
We present an experimental demonstration of a single-shot, non-destructive electron beam diagnostic based on the ionization of a low-density pulsed gas jet. In our study, 7~MeV electron bunches from a radio frequency (RF) photoinjector, carrying up to 100 pC of charge, traversed a localized distribution of nitrogen gas (N$_2$). The interaction of the electron bunches with the N$_2$ gas generated a correlated signature in the ionized particle distribution, which was spatially magnified using a series of electrostatic lenses and recorded with a micro-channel-plate detector. Various modalities, including point-to-point imaging and velocity mapping, are investigated. A temporal trace of the detector current enabled the identification of single- and double-ionization events. The characteristics of the ionization distribution, dependence on gas density, total bunch charge, and other parameters, are described. Approaches to scaling to higher electron bunch density and energy are suggested. Additionally, the instrument proves useful for comprehensive studies of the ionization process itself. 
\end{abstract}

\maketitle

\section{Introduction}

Advances in accelerator and source technology are propelling the generation of charged particle beams with unprecedented beam brightness, intensity and compactness in all spatial dimensions \cite{hooker2013developments, musumeci2018advances}. 
For example, the FACET-II facility can generate electron beams with energy of up to 10~GeV, and many kiloamp peak currents, focusable down to $<$ 20~$\mu$m spot sizes, ranking it among the highest peak intensity beams worldwide \cite{yakimenko2019facet}.
Conventional diagnostics are untenable at such high intensity, necessitating development of innovative techniques \cite{downer2018diagnostics}. Traditional transverse beam profile methods, in fact, such as optical transition radiation \cite{castellano}, scintillating screens \cite{walasek2012scintillating}, or wire scanners \cite{bazarov, orlandi2016design}, require insertion of solid-state density matter in the beam path. Thus, they have intrinsic limitations in handling very high beam intensity as they are susceptible to damage, especially near the beam focus. There is an urgent need for advanced, non-invasive techniques to extract beam parameters without intercepting the beam. Additionally, single shot acquisition adds substantial value because it would allow tagging each bunch in the accelerator for fast feedback, and to provide an online quality monitor for the final beam application. 

Recent progress in non-destructive beam diagnostic techniques has been focused on the optimization of speed, accuracy, and resolution. Among these, ionization-based beam monitors have been employed for many decades in high intensity accelerators, and particularly in ring machines where ionization of trace amounts of residual gas in the vacuum chamber provides a wealth of information on the circulating beams \cite{hornstra1967, hochadel1994}. Ionization-based profile monitors are by now an established technique in beam diagnostics at many particle accelerator laboratories, widely utilized especially in proton and ion beamlines and in order to characterize particle beams for medical applications \cite{wong2020introducing, thurman2023study,giacomini2004development}. In some cases, to boost the signal (proportional to the number of ions generated by the beam passage), a controlled amount of gas molecules can be injected timely in the vacuum to increase the target density, remaining orders of magnitude below solid-state density and therefore only minimally (ideally non) perturbing the main beam \cite{tzoganis2014, salehilashkajani2022gas, tzoganis2017design, zhang2023characterization}. 
Typically, the ionization products and the fluorescence light of the excited atoms/molecules are helpful in retrieving beam properties, including total charge, centroid position, and spot size \cite{yamada2021, shiltsev2021, scherkl}. 
Since photons are emitted in every direction by the gas molecules struck by the primary high-energy beam, the collection efficiency of the fluorescence signal is poor, and to achieve sufficient signal-to-noise ratio, the acquisition typically occurs on long integration times over multiple shots \cite{variola2007, salehilashkajani2022gas}. Conversely, the charged particles (both ions and electrons) from the ionization events can be redirected by a bias voltage, manipulated by optical elements, and collected using efficient detectors to maximize the retrievable information content. Combining a suitable injection scheme of gas molecules with the efficient detection of charged particles can significantly boost the technique sensitivity. This approach enables a novel implementation of ionization-based diagnostics where all the information can be retrieved after a single beam passage through the gas. 

This letter describes a non-invasive diagnostic technique for measuring transverse parameters from high-intensity, relativistic electron beams. 
The detection scheme is based on the passage of a relativistic electron beam through a low-density gas jet, which leaves a characteristic imprint in the from of an ionized distribution, that is subsequently measured to retrieve primary beam properties. The advantages of this diagnostic directly stem from the rich physics underlying the ionization process. In most cases, the electric field of the beam is not sufficiently high to ionize the gas molecule, and most of the ionization occurs via impact ionization. The number of ionization events is then directly correlated with the beam charge, and assuming the initial gas molecule distribution is much larger than the beam dimensions, the transverse ionization distribution is a faithful representation of the incoming beam profile \cite{tzoganis2014}. Under this condition, acceleration and transport of the ions through a high magnification imaging system enable the direct measurement of small-size features in the main beam transverse profile. On the other hand, when the beam space charge field is high enough (typically above 5 GV/m) to ionize the particles directly by tunneling through the potential barrier, the ionization profile is also strongly correlated with the beam charge density and peak current \cite{Tarkeshian2018}, and employing advanced algorithms allows the unfolding of the primary beam parameters from the detected ion distribution \cite{cook}. 

This study demonstrates the utilization of ionization-based beam diagnostics in a single-shot mode on a high brightness electron linac, which has not been previously reported. The experiment used the 7 MeV, 100 pC, ps-long high brightness electron bunches from the UCLA Pegasus RF photoinjector\cite{alesini2015new}.
A pulsed localized gas distribution with estimated peak density up to 10$^{14}$ cm$^{-3}$ was injected 5~m downstream of the photocathode, and synchronized to the arrival of the electron bunches. The number of generated ions was directly proportional to the total electron charge and the gas density. Time-of-flight traces indicate the occurrence of both single and double ionization events. Nearly order-of-magnitude magnification of the beam transverse profile was achieved with the ion imaging setup, an array of electrostatic lenses with independent remotely controllable voltages. In this regard, two modalities are investigated consisting of point-to-point imaging as well as a velocity mapping configuration \cite{eppink1997velocity} to visualize the initial transverse momentum distribution. Finally, the instrument versatility is demonstrated by switching the polarity of the imaging lenses for the detection of electrons, as opposed to ions. 

The paper is structured as follows: We discuss the experimental setup of the gas sheet monitor (GSM), which includes a detailed examination of the electrostatic lens column and the gas jet. Impact ionization is elucidated as the primary ionization pathway during this experiment, accompanied by estimates for ion yield. Key metrics, such as time-of-flight and centroid tracking, align excellently with the theoretical predictions. The main findings highlight the effectiveness of the GSM in a single-shot modality for characterizing the primary electron beam distribution centroid and standard deviation through the analysis of ionization images. Finally, we conclude by discussing the potential applicability and extensions of this diagnostic technique.

\section{Experimental Setup}
The GSM commissioning experiment was conducted at the UCLA Pegasus high-brightness electron beam facility. Before deployment on the electron beam line, the GSM subsystems were tested and optimized on a tabletop using ultrafast laser ionization of nitrogen gas to characterize each of the components, and benchmark the charged particle transport model with particle tracking simulations. While the setup can be used with different gaseous species, nitrogen gas was chosen for these initial experiments as it is most effectively pumped by the turbomolecular pumps in order to reduce the load on the ultra-high vacuum system.  The gas delivery, high voltage electrostatic column, and imaging layout were evaluated and operational ranges were established. The complete diagnostic was then installed on the Pegasus electron beam line, for use with transversely focused high-charge electron bunches.

\begin{figure}
    \centering
\includegraphics[width=0.5\textwidth]{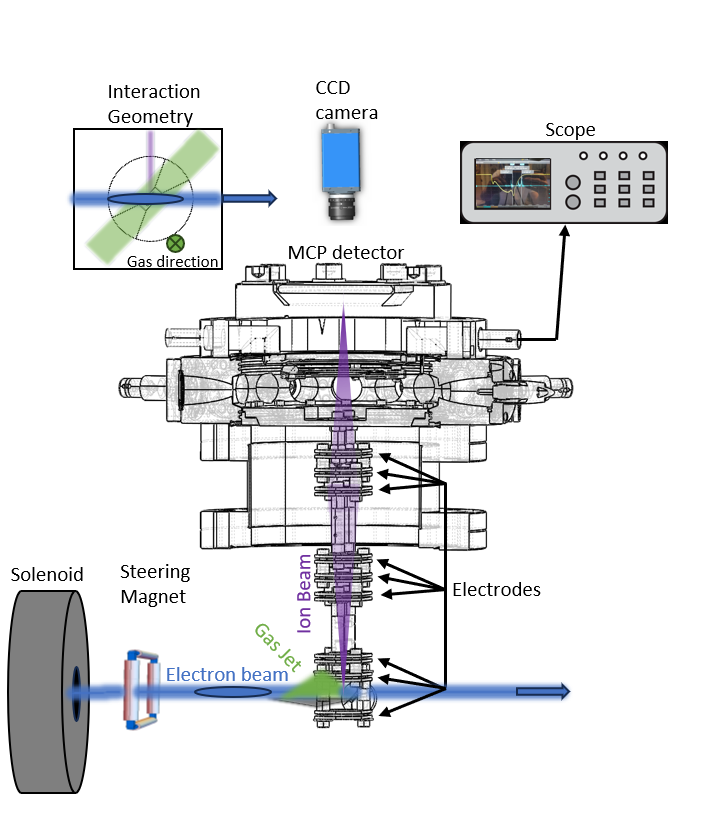}
    \caption{Sketch of the gas sheet monitor (GSM) layout. A solenoid and a steering magnet are used to focus and position the relativistic electron beam (blue line) from the Pegasus photoinjector into the gas jet (green) where the ionization occurs. The trajectories of  molecular nitrogen ions through the column are shown in purple. The ion column electrode voltages are tuned to direct the charged particles to the micro-channel plate (MCP) detector where an image of their spatial distribution is captured using a CCD camera. An oscilloscope is used to monitor the ion current and time-of-flight.}
    \label{fig:setup}
\end{figure} 
\begin{figure*}
    \centering
        \includegraphics[width=0.35\textwidth]{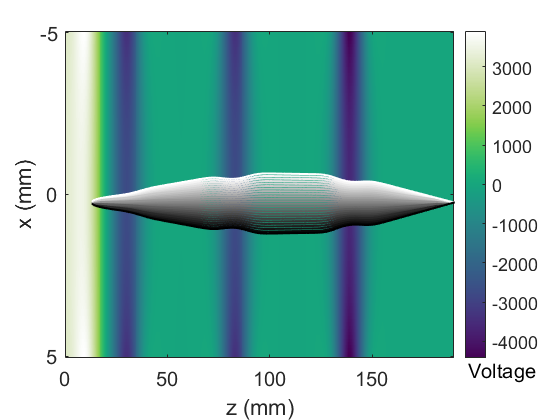}(a)
        \includegraphics[width=0.35\textwidth]{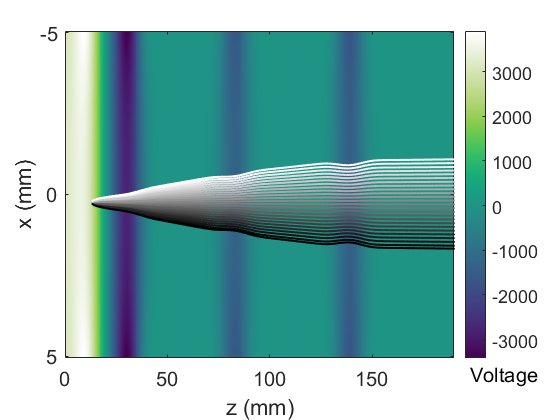}(b)
    \caption{The trajectories of molecular nitrogen  ions simulated using the GPT particle tracking code through the column are superimposed with the corresponding electrostatic potential generated by the electrodes. The ion trajectories are color coded in a gray scale based on their initial angle ($\pm$ 100 mrad). (a) The electrostatic column voltages are adjusted for imaging to capture the initial ion transverse spatial distribution. (b) The voltage distribution is configured to map the initial velocity, or angle, information to the detector while minimizing the dependence on the initial position.}
    \label{fig:traj}
    \label{fig:traj}
\end{figure*} 

The experimental setup at the UCLA Pegasus beamline is described here. A Ti:sapphire laser system was used to generate laser pulses with a central wavelength of 780~nm and pulse duration of 40~fs. The laser was converted to UV using non linear crystals, stretched using birefringent crystals and then directed to an NaKSb cathode located inside a S-band RF photogun, where electron bunches were generated with upwards of 100~pC of total charge. 
In the photogun section, a peak accelerating gradient of 75~MV/m was initially utilized to accelerate the electron beam to 3.2~MeV kinetic energy. Then, the first solenoid focused the beam into a booster linac which further accelerates the particles to reach 7~MeV kinetic energy. A second solenoid focused the beam into the interaction chamber transversely, and a steering magnet was positioned just after the final solenoid. The steering magnet was used to adjust the point of interaction between the electron beam and the gas jet within the interaction chamber. Differential pumping was employed to maintain the gun pressure at or below $10^{-9}$ Torr as necessitated by alkali-antimonide photocathode operation.

A schematic showing the main elements of the GSM is depicted in Fig. \ref{fig:setup}. The relativistic electron beam collides with the nitrogen gas emanating from the $45^{\circ}$ tilted skimmer, generating both ions and secondary electrons. 
Depending on the electrostatic column polarity, the ions or secondary electrons can be transported to form an image on a microchannel plate (MCP). 
The MCP employed in the experiment is a two-stage device. 
The MCP is operated in a voltage range of 1.4 to 2~kV in the experiment, which corresponds to a proportional gain ranging from 50 to 60~dB. 
The homogeneity of the response over the detector area was not studied in detail, but it should be noted that only a relatively small area in the center of the MCP was used for these studies.
The voltage on the electrostatic rings, or lenses, along the column are controllable to allow different magnifications for imaging, or to operate in velocity mapping mode. 

The electrostatic column comprises nine conducting rings of 7.9~mm radius, each with adjustable potential for distinct functionality depending on the mode of operation. 
The first two rings (repeller and extractor) are located on either side of the interaction point and are tuned at equal and opposite potentials to impart an initial kinetic energy to the ionized particles. The third ring is kept at a potential one tenth of the repeller and serves as a focusing electrode to control the initial beam divergence. Altogether, the three first rings serve to accelerate and collimate the initial ion distribution. The second and third triplets in the column act as electrostatic Einzel lenses \cite{adams1972}.  In this configuration, for each lens the central ring is kept at a potential ten times greater than the two adjacent rings, albeit with opposing polarity.

Charged particle transport through the system was simulated using the General Particle Tracer software (GPT \cite{gpt:pulsar}), which tracks particles through a cylindrically symmetric field map generated using Poisson \cite{menzel1987users}. We used this model to determine the optimal voltage configurations for achieving real space or velocity map imaging of the nitrogen ions or the secondarily produced electrons generated during the ionization process. The transverse position of an ion at the final image plane, particularly in the horizontal plane, is determined by the initial conditions $(x_0,p_{x0})$. In the linear optics approximation, this can be written as $x=R_{11}x_0+R_{12}p_{x0}/mc$, where $R_{11}$ and $R_{12}$ are the matrix elements of the transport system, $m$ is the mass of the ion, and $c$ is the speed of light. Real space imaging occurs when the final image is independent of the initial velocity of all particles, or $R_{12}=0$.  Alternatively, velocity map imaging occurs when the final image is independent of the initial position, i.e., $R_{11}=0$. 

Fig. \ref{fig:traj} illustrates two of the numerous potential configurations resulting from GPT simulation results, demonstrating (a) imaging and (b) velocity-map-imaging (VMI).
In each case, the potential distribution is displayed as a contour plot in the background behind the trajectories. In the transverse direction the electric potential varies by less than 0.5 \% over a distance of 5 mm from the axis of the column. Particle trajectories are color-coded from dark to light grey to indicate correlations with the initial angle. In (a), during imaging, a bundle of rays emanating from a point is imaged back to a point in the detector plane. In (b), during VMI, a bundle of rays fans out according to the initial angle. The repeller/extractor rings were held at 5 kV in both scenarios. For imaging, the central rings in the Einzel lenses were tuned to 5 kV and 6.5 kV, respectively. For VMI, the first and second Einzel lenses were tuned to 5 kV and 2.5 kV, respectively.

During the experiment, the nitrogen gas reservoir was maintained at atmospheric pressure and room temperature. A piezo-actuated nozzle with a diameter of 0.2 mm, positioned 102 mm away from the interaction point, was used to pulse the gas injection into the chamber at a maximum 1 Hz repetition rate, mainly limited by the capacity of the vacuum turbomolecular pumps on the beamline to absorb the gas load and maintain an operating pressure in the 10$^{-6}$ torr range, although peak pressure briefly reached close to 10$^{-5}$ torr during gas pulses. The pulse duration could be varied between 50 to 250 $\mu$s. A slit aperture with dimensions $a$=1~mm and $b$=3~mm was placed near the interaction point to collimate and shape the gas distribution for interaction.

It is possible to estimate the density of gas molecules at the interaction point based on the reservoir temperature, pressure, and geometry. Assuming a thermal velocity distribution and a small solid angle, the gas flux at the interaction point can be written as:
\begin{equation}
\phi = n_{res} A_1\int v_x f(v_x,v_y,v_z) d^3v \approx \frac{1}{\sqrt{6\pi}}\frac{A_1}{\pi D^2}n_{res} A_2 v_{p}
\label{eq:lamflow}
\end{equation}
where $A_1$ and $A_2$ represent the area of the nozzle and the collimator respectively, $D$ is the distance between them, $n_{res}=P/kT$ is the gas density in the reservoir at near standard pressure $P$ and temperature $T$, $k$ is Boltzmann constant, $v_{p} = \sqrt{2kT/m_{N2}}$ is the most probable speed, $m_{N2}$ is the molecular mass of nitrogen. We can then estimate the interaction point density as $n \approx \phi/v_{p}A_2$, which yields for peak gas number density (i.e. number of molecules per unit volume) at the interaction point a range of  $10^{13}$ and $10^{14}$ cm$^{-3}$ in agreement with measurements performed with a localized pressure gauge sensor during the preliminary tests of the chamber \cite{burger2022experimental, zhang2023characterization}. The calculation in Eq. \ref{eq:lamflow} assume laminar flow, which is justified since for the number density above, a dynamics viscosity of $\mu$=1.76$x10^{-5}$ Pa sec and a characteristic length of 1 mm, we can calculate a Reynolds number well below the typical threshold of 2000 for turbulent flow.

\begin{figure*}[ht]
    \centering   \includegraphics[width=0.305\textwidth]{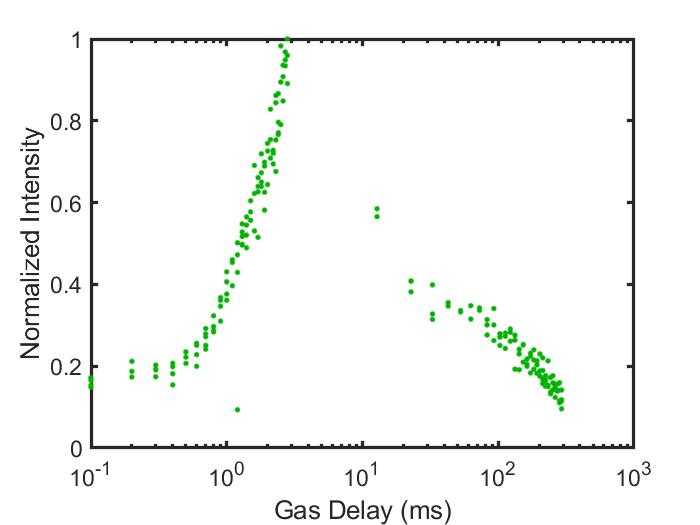}(a)
\includegraphics[width=0.305\textwidth]{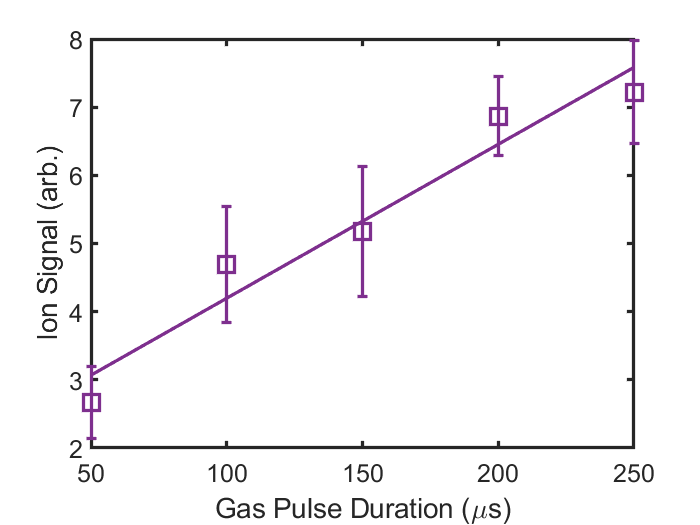}(b)
    \caption{(a) Measurements of ionization intensity as a function of the relative delay between the opening of the gas valve and the e-beam arrival time. (b) Ion signal intensity measured as a function of the gas pulse duration. Note that the background signal originating from the electrical noise on the MCP has not been subtracted in these images and is responsible for the baseline levels in the normalized intensity.}
    \label{fig:timing}
\end{figure*} 

Next, the yield of N$_2^+$ ions can be estimated using the average energy loss rate by an electron when traversing the nitrogen gas column, which for an electron of rest mass $mc^2$, relativistic factor $\gamma$, and relative velocity $\beta = \sqrt{1-1/\gamma^2}$ can be written as:
\begin{equation}
    \frac{dE}{ds} = 4\pi r_c^2 n Z  \frac{mc^2}{\beta^2}\left(\ln(B_q) + f(\gamma) \right)
\end{equation}
where  $r_c$ is the classical electron radius, $n$ is the number of gas molecules per unit volume, $Z$ is the number of electrons in the nitrogen molecule, $B_q=\gamma\beta m c^2 \sqrt{\gamma-1}/I_\omega$, wherein, $I_{\omega}(eV)\approx(9.76 + 58.8Z^{-1.19})Z$ is an expression used to approximate the effective excitation energy, and finally the term $f(\gamma) = \left[\frac{(\gamma-1)^2}{8} + 1 - (\gamma^2 + 2\gamma - 1)\ln(2)\right]/2\gamma^2$, which approaches a constant value at higher energy   \cite{Tsoulfanidis1995MeasurementAD}.

For an incident electron bunch with total charge $Q_b$, assuming most of the energy loss goes into ionization, we can estimate the ion charge produced with the expression:
\begin{equation}
    Q_I\approx Q_b\frac{dE}{ds}\frac{\Delta s}{\epsilon} \label{eq:yield}
\end{equation}
where $\epsilon = 15.58$~eV is the ionization potential of molecular nitrogen, and $\Delta s=\sqrt{2\pi}a/\cos(45^{\circ})$ is the effective thickness of the gas jet near the interaction. According to Eq. \ref{eq:yield}, we expect between 1~fC and 10~fC of ion charge yield for incident beam charges ranging from 10 to 100 pC. The number of ionization events can also be calculated by extrapolating to relativistic energies \cite{kim2000extension} the NIST cross section for the ionization of molecular nitrogen by electrons yielding $\sigma$ = 10$^{-22}$ m$^2$ \cite{kim1994binary,kim2010electron}. The number of ions generated per incident electrons can then be estimated as $\sigma n \Delta s$ and is found to be consistent with the result from Eq. \ref{eq:yield}.
\begin{figure*}[ht]
    \centering
        \includegraphics[width=0.305\textwidth]{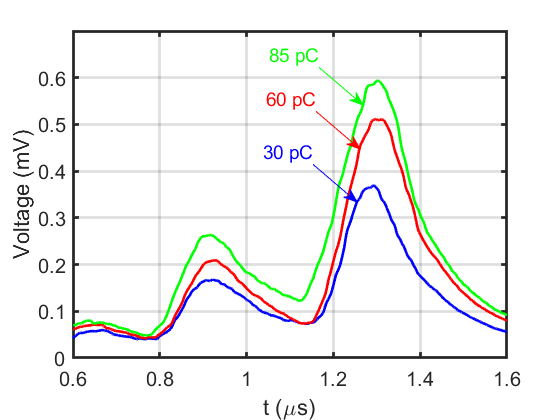}(a)
        \includegraphics[width=0.305\textwidth]{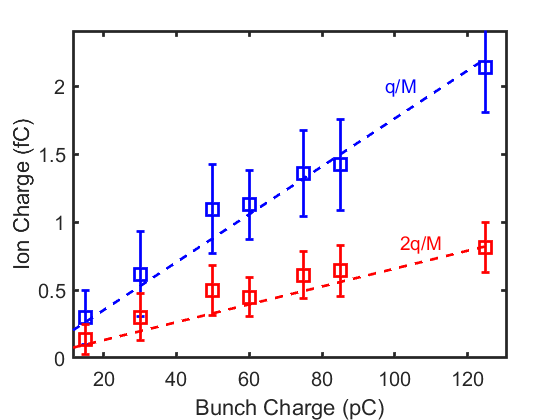}(b)
        \raisebox{-0.25cm}{\includegraphics[width=0.275\textwidth]{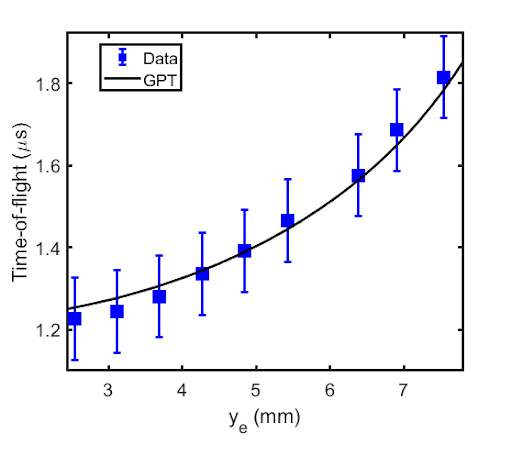}}(c)
    \caption{(a) Time-of-flight distribution of the ions collected by the MCP measured on a scope for different bunch charges. (b) Collected ion charge in each of the two main peaks of the oscilloscope trace as a function of the primary beam charge. (c) Time delay of the main N2+ peak as a function of the position of the relativistic electron beam between the first two electrodes is varied using the upstream steering magnet. The fitted curve from GPT simulation allows us to determine the initial vertical position of the ions in the column, which is used to offset the x-axis tick labels.}
    
    \label{fig:signal}
\end{figure*}
\begin{figure*}
    \centering       
    \includegraphics[width=0.355\textwidth]{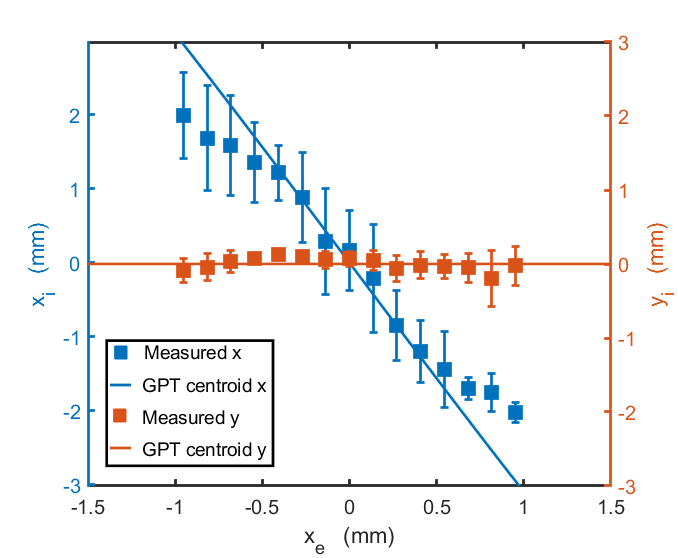}(a)
\includegraphics[width=0.355\textwidth]{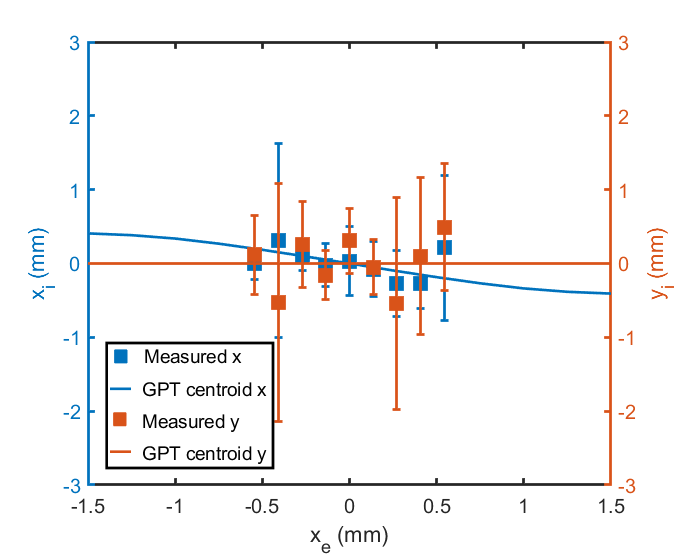}(b)
    \caption{Measurement of the horizontal and vertical centroid positions ($x_i$ and $y_i$ respectively) of the ionization distribution on the MCP as a function of the horizontal position of the primary beam ($x_e$) acquired while the electrostatic column was set to imaging (a) and velocity mapping (b) conditions.}
    \label{fig:Control}
\end{figure*}

\section{Experimental Results}

During the experiment, two signals were simultaneously acquired from the setup: an intensified image of the transverse ion distributions captured by a CCD camera observing the MCP, and an oscilloscope trace illustrating the time-of-arrival distribution of ions on the MCP. The latter was obtained by connecting the input electrode of the MCP to a digital oscilloscope.

To establish synchronization between the arrival times of the gas jet and the electron beam, we performed a scan of the gas pulse delay. A notably enhanced ionization signal was observed on the CCD camera when the electrons’ arrival time aligned with the gas valve’s opening. The peak ionization occurred with a 2 ms delay between the arrival of the electron beam and the trigger to the gas piezo valve.  The rise time in Fig. 3a can be understood as due to the spread in the time of arrival of the gas molecule to the beam axis. The ionization signal did not significantly diminish for long delays, which may be attributed to a persistent gas trail over approximately 100 ms. This temporal behavior is depicted in Figure \ref{fig:timing}(a).  

Increasing the gas pulse duration resulted in a linear increase in the background pressure and consequently a linear gain in the ionization signal as shown in Figure \ref{fig:timing}(b). The error bars in the figure reflect charge fluctuations measured at the MCP and are calculated as the standard deviation of the intensity distribution over 5 images. These charge fluctuations are mainly due to the primary beam charge variations as well as the MCP gain fluctuations. For this set of experiments, we operated with a gas pulse duration of 200~$\mu$s as this provided a signal-to-noise sufficient to determine the beam transverse size in a single shot at a level comparable with the shot-to-shot fluctuations in the beamline, while at the same time keeping the gas load on the beamline manageable.

The oscilloscope waveform can be converted to an absolute time-of-flight measurement by referencing the photo-excitation laser arrival to the cathode as time zero, since the relativistic electron beam takes just $<$ 10 ns to travel down the beamline. 
Fig. \ref{fig:signal}(a) shows typical waveforms recorded for different input beam charges on the scope. The traces present a main peak at 1.3 $\mu$s consistent with the time-of-flight of N2+ ions from the interaction point to the MCP plane through the ion column. A second peak is observable at 0.9 $\mu$s formed by ions a with charge to mass ratios differing by a factor of 2. A third much smaller peak at 0.65 $\mu$s which would correspond to a charge to mass ratio of 4 could also be distinguished. 
It is worth noting that while the heights of the peaks varied in relation to changes made to the electron beam focusing and steering, their relative ratio and time separations remained constant. 

We attributed the different peaks to different impact ionization channels for the nitrogen molecule after collision with a relativistic electron. In principle relativistic electrons can easily lose multiple tens of eV in the collision, opening up a significant number of decay possibilities for ionization, including the creation of N$_2^{2+}$ ions, or the break up of the molecule into N or N$^{+}$ ions. 

The ionization threshold for the double ionization or direct breaking of the molecule into  two N+ ions are above 40 eV\cite{dutuit2013critical}. This will make it significantly more unlikely for this kind of events to occur. Note that in our diagnostic we can not distinguish if the second peak in the waveform is due to N$_2^{2+}$ or to fragmentation in N$^{+}$ ions since these particles have the same charge-to mass ratio. Additionally, since the scope signal adds up the collected charge, the second peak will be twice higher for the same number of primary electron collisions, leading to an estimate for the cross section ratio with respect to the single ionization channel of approximately 4.5:1 in agreement with the expectations from the extrapolation of cross-section data for low energy electrons reported in \cite{itikawa1986,rapp1965}. The third small peak matches N$^{2+}$ ions which can be generated when an even larger amount of energy (more than 50 eV) are exchanged in the collision. It is worth noting that in the table-top laser tests of the setup, only the main peak corresponding to N$_2^{+}$ ions could be seen, as the impact ionization mechanism is fundamentally different than laser-field ionization where for a given incident intensity, one finds exponentially decreasing probability for higher ionization states.

In order to retrieve an absolute value for the ion current, the oscilloscope was pre-calibrated prior to installation on the beamline by using a femtosecond laser and a picoamp meter. A linear relation between the waveform peak amplitude and the current read out was established. Taking into account the repetition rate of the laser, the current was converted into a charge per pulse. The integrated intensity on the MCP image was also found to be proportional to the charge in the ion beams, but in this case the constant of proportionality is dependent on the gain (and therefore on the voltage settings) of the MCP. 

The beam charge incident onto the gas jet was then varied to measure the proportional change in the ionization yield. The scan results are shown in Fig. \ref{fig:signal} (b), where the total charge of each population of ions is presented. Note how the ratio of the two populations remains fairly constant as the bunch charge is changed over an order of magnitude up to 100 pC ruling out the possibility that higher order ionization states are due to multiple impact ionization events. 

The upstream steering magnet was then used to benchmark the lens voltage setpoints and demonstrate that the electrostatic column could be operated in imaging and velocity mapping mode. The steering magnet had been previously calibrated to relate the power supply current to the horizontal and vertical kicks in the electron beam trajectory. Vertical steering effectively shifts the initial plane in the column where the ion distribution starts in the potential, directly influencing the time-of-flight to the MCP detector. 

A measurement of the position of the main peak in the oscilloscope waveform as the electron beam was vertically steered is shown in Fig. \ref{fig:signal}(c). A fit of the data to the GPT transport model allows to pinpoint the location where the ions are being generated in the potential between the first two electrodes. The comparison of the time-of-flight measured to the particle tracking model prediction is shown in Fig. \ref{fig:signal}(c). Note that using the results of the fit the horizontal axis in this plot has been used to offset the initial vertical position of the electron beam in the column $y_e$ above the first electrode. 
This approach ultimately served as an important benchmark of the particle tracking model, and provided a key-input to the simulation (initial ion position) to tune the lenses voltage setpoints and obtain an imaging condition at the MCP detector.

Horizontally steering the electron beam and then tracking the corresponding position of the centroid of the ion beam on the MCP yields a measure of how the final image is correlated with the initial ion distribution. 
In Fig. \ref{fig:Control}(a) we show the results of the scan with the voltages in the column tuned to the imaging configuration. 
As expected, the ions mainly moved along one direction on the MCP screen owing to the column alignment which minimizes the x-y coupling in the transport. A linear fit of the data yields the transport coefficient $R_{11} = 3.2$, which represents the magnification of the column, in agreement with the GPT model prediction. Some small non-linearities which can be attributed to the aberrations in the electrostatic lenses, are also observed for large offsets in the electron beam horizontal position. Note that one challenge in performing this scan is that steering the electron beam into different regions of the gas jet profile changes the amplitude of the signal. Based on the geometry of the gas delivery the region that generates the most ions is as close to the nozzle as the beam can be, but during operation care must be taken to avoid collisions of the primary beam with the gas nozzle, which can lead to unwanted ionization events and, subsequently, charging of the electrostatic column. Ultimately, after retuning the voltage setpoints on the electrostatic lenses, we also demonstrated our ability to minimize $R_{11}$ and establish a velocity mapping imaging condition (with an estimated $R_{12}$
= -66 m). The results of this scan is  presented in Fig. \ref{fig:Control}(b) where the horizontal position of the ion beam on the MCP is shown to be essentially independent of the primary electron beam horizontal position. 

\begin{figure}
    \centering
    
\includegraphics[width=0.45\textwidth]{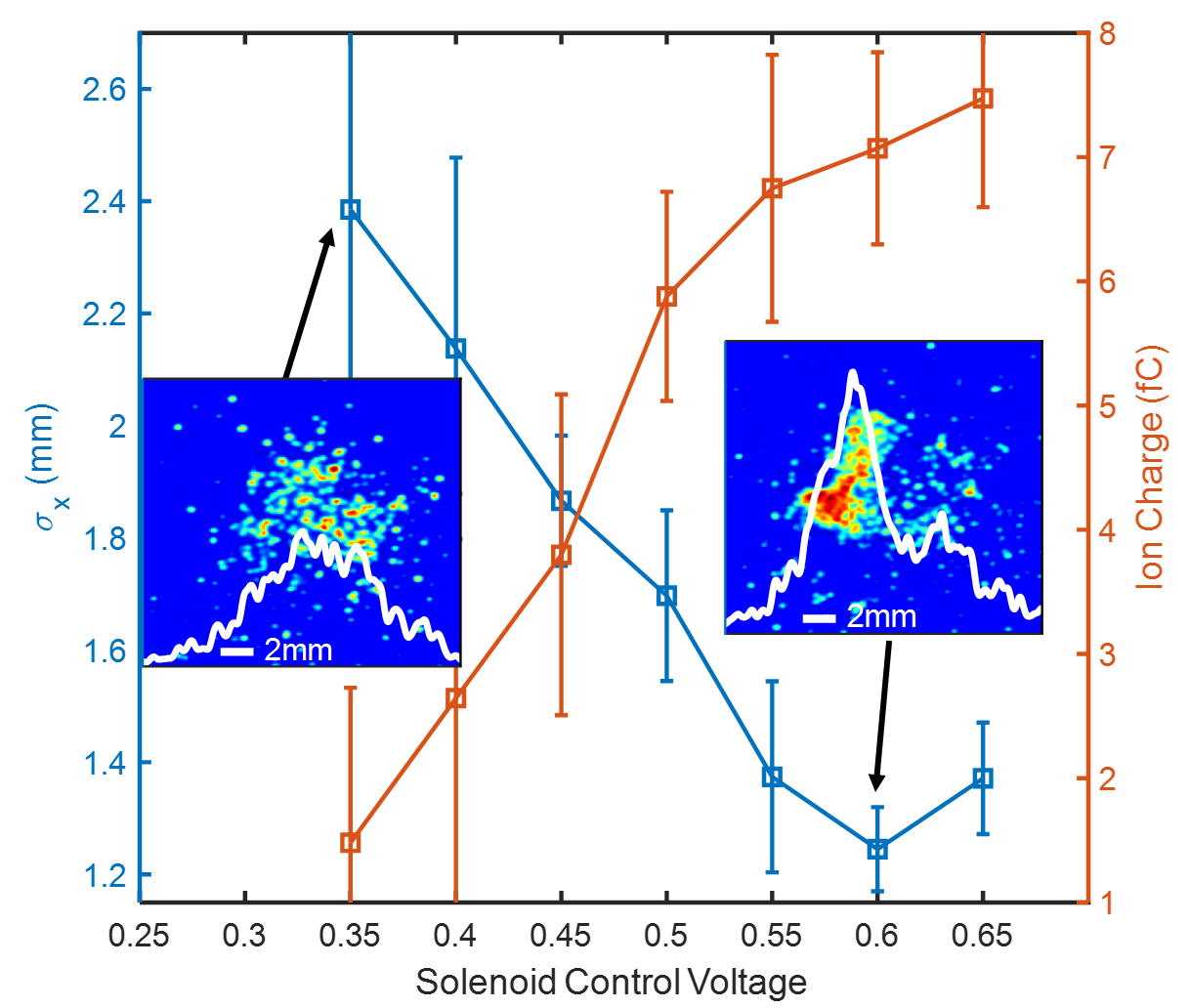}(a)
\includegraphics[width=0.45\textwidth]{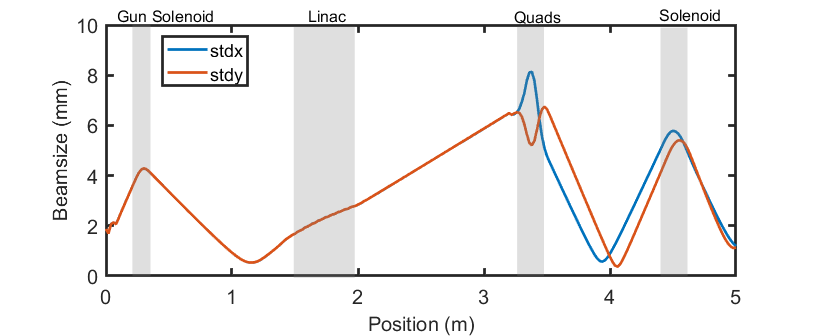}(c)
    \caption{(a) Horizontal rms size of the ion distribution measured on the MCP as the electron beam is focused onto the gas jet by the solenoid. The right axis shows a correlated increase in ion charge as the solenoid current was increased. The left inset shows an image of the ion distribution captured on the MCP when the solenoid was underfocused.  The right inset shows an image of the ion distribution at the beam waist. (b) Transverse beam size evolution from a start-to-end simulation of the Pegasus injector. Experimental settings achieve a 1.19 mm beam waist at the interaction point.} 
    \label{fig:solenoidscan}
\end{figure} 
While in the imaging configuration with a magnification of 3x, we also performed a solenoid scan with the bunch charge set at 60~pC by varying the electron beam size at the point of interaction and collecting the corresponding images of the ion distribution at the MCP. The images are analyzed to retrieve the horizontal rms size and signal intensity of the ion distribution as a function of the solenoid strength. The minimum ion distribution size and peak signal are obtained for the solenoid setting where the electron beam waist is located in the middle of the GSM chamber. The analysis results are shown in Fig. \ref{fig:solenoidscan}(a). In the inset, we also show the raw MCP images of the ions when the electron beam is under-focused and at the waist, respectively, along with their respective horizontal projections.
The retrieved values for the rms beam sizes at the GSM location are consistent with the results from GPT particle tracking simulations of the Pegasus beam line at the gas jet plane. 
In Fig. \ref{fig:solenoidscan}(b) we show start-to-end simulations for the evolution of the rms beam size along the beam line for a 60~pC beam charge, including the final solenoid which focuses the beam into the GSM chamber positioned at 5~m. 
The final rms sizes of the beam waist are 1.19~mm and 1.13~mm, in the horizontal and vertical dimensions respectively, in good agreement with the spot size retrieved by the GSM diagnostic. 
Note that the finite size of the gas cloud sets a limit on the largest beam size that can be measured.


The increase in the ion signal through the solenoid scan can be attributed to the finite dimensions of the gas distribution in the chamber. As the electron beam size becomes comparable with the size of the gas cloud, in fact, more electrons sample a higher density region of the gas distribution and the number of ions increases. Taking into account the magnification of the column we can estimate the spatial extent of the gas jet to be close to 2~mm in excellent agreement with the estimate from the nozzle geometry. 



\section{Conclusion and outlook}

In conclusion, we demonstrated a novel experimental apparatus for the single-shot diagnostic of electron beams based on monitoring the ionization signal from the passage of the beam in a gas jet.
Properly tuning the voltages in a multilens electrostatic microscope column, we were able to transport the charged particles generated in the ionization process to a microchannel plate detector both in imaging and velocity mapping mode. Operating in imaging mode, the setup enabled the measurement of the transverse extent of the beam with a resolution limited by the point spread function of the detector which can be estimated from the single ion hits to be 0.19 mm rms. Taking into account the measured magnification of the column, the rms spatial resolution in these first experiments was below 0.07 mm. Further improvements in the resolution could be obtained by increasing the column magnification. The gas density of the thin gas curtain generated by the nozzle was below 10$^{15}$~cm$^{-3}$, so only an extremely small fraction of the beam is affected in the interaction, making this non-destructive diagnostic ideally suitable for monitoring very intense beams. 

In the experiment, a linear relationship between bunch charge and ionization signal was observed during synchronization indicating that impact ionization was the main process at play in the interaction of the electrons with the gas consistent with the beam charge and beam dimensions employed in the tests. It is envisioned that when more extreme beam parameters are used (higher charge, shorter bunch lengths, smaller spot sizes), field ionization would strongly contribute to the number of generated ions, which would correlate with the beam peak current indirectly yielding the temporal bunch length and further extending the information that could be retrieved from the apparatus. 

During operation, a primary challenge arose in charging the electrostatic column.
This issue was typically encountered during the initial establishing of the signal, and led to a distortion of the ion beam as it deviated from its intended axis. For next version implementations, incorporating a bipolar power supply which would allow a fast turn-around column discharge would be advantageous to remedy the charging. Note that by switching the sign in the accelerating electrodes, it is also possible to reconfigure the column to collect on the MCP the secondary electrons produced during the impact ionization process. Both velocity mapping and imaging mode for the electrons were verified during the experiment even though no quantitative data was acquired. In order to  image the electrons,  the required voltages in the column were found to be 2.5kV for the repeller/extractor, and 2.5kV in each einzel lens. VMI required 1.75kV for the repeller/extractor and 2.5kV, and 2.85kV for the first and second Einzel lens respectively. 

The electrons could also be used to retrieve information about the incident electron beam, however, extracting beam parameters requires more care as due to the large mass difference the electrons are more susceptible to their collective space charge forces. Secondary electron pulses generated from ionization are in fact more prone to blurring due to their own space charge field during transport, unlike ions, which are negligibly influenced by their own space charge field. Note that both ion and electron imaging are affected by the primary beam space charge field, leading to additional magnification which has been theoretically investigated \cite{SHILTSEV2021164744}. This additional space-charge expansion factor depends only on the primary beam parameters and the extracting field but not on the secondary mass or charge. For specific parameters relevant to the PEGASUS beamline (e.g., a 60 pC, 10 ps beam, with a 500 kV/m extraction field applied over a 3 mm distance), this contribution to the transport is estimated to be negligible. This is supported by start-to-end simulations that show that primary beam space charge effects in the Pegasus setup  are perturbative at best, since the impulse on ions from the primary beam electric field being much smaller than that from the extraction field in the time it takes the primary beam to pass by. The VMI modality could potentially be used to visualize the initial momentum distribution of the secondary electrons. In combination with a short laser pulse this would have an imprint on the ionization time as well as the primary beam space charge field \cite{siqi}.

There is ample room for future modifications in the experimental setup, including precision collimation of the neutral molecular beam to further localize the initial ion distributions, and an adjustment of the microscope column, aided by optimizing particle tracking simulations, to increase the magnification in imaging mode and enhance spatial resolution. Increasing the gas density would further enhance the signal by generating more ions. This would facilitate the detection of smoother images and projections like those in existing literature \cite{HASHIMOTO2004289,PhysRevAccelBeams.24.042801}. By increasing ion charge, higher extraction fields may be needed to mitigate space charge effects. In our setup, the vacuum system limited the gas density, which would require upgrades to handle the higher load. Future improvements to the pumping configuration will allow for greater gas density while maintaining system performance. Despite this, increasing the gas density by an order of magnitude is expected to keep the instrument non-invasive or minimally invasive, as the energy loss within the small interaction region remains small compared to the beam's initial energy spread. For larger number of ionization events, it will become necessary to consider other aspects such space charge effects in the microscope column transport, as well as the effects of the primary beam field on the initial angular distribution, and aberrations due to nonlinearities in the transport. 

Time-of-flight measurements highlighted the generation of multiple ion species from the interaction of the relativistic electrons with molecular nitrogen gas. In fact, the different ion species follow different evolution in the column so that imaging and velocity map conditions can not possibly apply to all the ionization products, thus blurring the desired signal for the diagnostics. Due to the relatively large temporal separation in the time-of-arrival of the different species, a short high voltage gate pulse on the MCP electrodes could be used to select only one ion species and sharpen the image. However, this effect also highlights the versatility of the instrument, as the richness of the ionization spectrum suggests the possibility to use the system as a basic scientific instrument to characterize ionization cross sections at relativistic energies, which is an area of interest across many disciplines such as astrophysics and plasma physics. 


\newpage

\newpage
\begin{acknowledgments}

This work supported by DOE grant No. DE-SC0009914 and SBIR contracts No. DE-SC0019717 and DE-SC0017691. This work was also in part supported by the National Science Foundation under Grant No. DMR-1548924.

\dots.
\end{acknowledgments}

\appendix

\bibliography{references}

\end{document}